\documentclass[letter,traditabstract]{aa}
\usepackage{txfonts}
\usepackage{graphicx}
\begin{document}


\title{{\it Herschel} photometry of brightest cluster galaxies in cooling flow clusters 
\thanks{{\it Herschel} is an ESA space observatory with science instruments provided by European-led Principal Investigator consortia and with 
important participation from NASA.}}

\author{A. C. Edge\inst{\ref{inst1}} \and 
J. B. R. Oonk\inst{\ref{inst2}} \and 
R. Mittal\inst{\ref{inst3}} \and
S. W. Allen\inst{\ref{inst4}} \and
S. A. Baum\inst{\ref{inst3}} \and
H. B\"ohringer\inst{\ref{inst5}} \and
J. N. Bregman\inst{\ref{inst6}} \and
M. N. Bremer\inst{\ref{inst7}} \and
F. Combes\inst{\ref{inst8}} \and
C. S. Crawford\inst{\ref{inst9}} \and
M. Donahue\inst{\ref{inst10}} \and
E. Egami\inst{\ref{inst11}} \and
A. C. Fabian\inst{\ref{inst9}} \and
G. J. Ferland\inst{\ref{inst12}} \and
S. L. Hamer\inst{\ref{inst1}} \and
N. A. Hatch\inst{\ref{inst13}} \and
W. Jaffe\inst{\ref{inst2}} \and
R. M. Johnstone\inst{\ref{inst9}} \and
B. R. McNamara\inst{\ref{inst14}} \and
C. P. O'Dea\inst{\ref{inst15}} \and
P. Popesso\inst{\ref{inst5}} \and
A. C. Quillen\inst{\ref{inst16}} \and
P. Salom\'e\inst{\ref{inst8}} \and
C. L. Sarazin\inst{\ref{inst17}} \and
G. M. Voit\inst{\ref{inst10}} \and
R. J. Wilman\inst{\ref{inst18}} \and
M. W. Wise\inst{\ref{inst19}}
}

\institute{ Institute for Computational Cosmology, Department of Physics, Durham University, Durham, DH1 3LE, UK \label{inst1} \and
 Leiden Observatory, Leiden University, P.B. 9513, Leiden 2300 RA, The Netherlands \label{inst2} \and
 Chester F. Carlson Center for Imaging Science, Rochester Institute of Technology, Rochester, NY 14623, USA \label{inst3} \and
 Kavli Institute for Particle Astrophysics and Cosmology, Stanford University, 452 Lomita Mall, Stanford, CA 94305-4085, USA \label{inst4} \and
 Max-Planck-Institut f\"ur extraterrestrische Physik, 85748 Garching, Germany\label{inst5} \and
 University of Michigan, Dept. of Astronomy, Ann Arbor, MI 48109, USA \label{inst6} \and
 H H Wills Physics Laboratory, Tyndall Avenue, Bristol BS8 1TL, UK\label{inst7} \and
 Observatoire de Paris, LERMA, CNRS, 61 Av. de l'Observatoire, 75014 Paris, France\label{inst8} \and
 Institute of Astronomy, Madingley Rd., Cambridge, CB3 0HA, UK\label{inst9} \and
 Michigan State University, Physics and Astronomy Dept., East Lansing, MI 48824-2320, USA\label{inst10} \and
 Steward Observatory, University of Arizona, 933 N. Cherry Avenue, Tucson, AZ 85721, USA\label{inst11} \and
 Department of Physics, University of Kentucky, Lexington KY 40506 USA\label{inst12} \and
 School of Physics and Astronomy, University of Nottingham, University Park, Nottingham NG7 2RD, UK\label{inst13} \and
 Department of Physics \& Astronomy, University of Waterloo, 200 University Avenue West, Waterloo, Ontario, Canada N2L 3G1 \label{inst14} \and
 Department of Physics, Rochester Institute of Technology, 84 Lomb Memorial Drive, Rochester, NY 14623-5603, USA \label{inst15} \and
 Department of Physics and Astronomy, University of Rochester, Rochester, NY 14627, USA\label{inst16} \and
 Department of Astronomy, University of Virginia, P.O. Box 400325, Charlottesville, VA 22904-4325, USA\label{inst17} \and
 School of Physics, University of Melbourne, Victoria 3010, Australia\label{inst18} \and
 ASTRON, Netherlands Institute for Radio Astronomy,P.O. Box 2, 7990 AA Dwingeloo, The Netherlands\label{inst19}
}

\abstract{The dust destruction timescales in the cores
of clusters of galaxies are relatively short given their
high central gas densities.
However, substantial mid-infrared
and sub-mm emission has been detected in many brightest
cluster galaxies.
In this letter we present {\it Herschel} \rm PACS and SPIRE
photometry of the brightest cluster galaxy in three strong cooling flow
clusters, A1068, A2597 and Zw3146. This photometry indicates that a substantial
mass of cold dust is present ($>3\times 10^7$~M$_\odot$) at temperatures
significantly lower (20--28~K) than previously thought based on
limited MIR and/or sub-mm results. The mass and temperature
of the dust appear to match those of the cold gas traced by CO 
with a gas-to-dust ratio of 80--120.}

\date{Received 30 March 2010/Accepted}


\maketitle

\section{Introduction}

The cores of cluster of galaxies are very energetic regions with a
high X-ray emissivity, particle density, cosmic ray flux, stellar density and AGN
radiation. In this very hostile environment any dust grains
are unlikely to survive for more than a few million years
due to the action of collisional sputtering (Dwek \& Arendt 1992)
unless they are shielded (Fabian et al. 1994).
It is therefore somewhat surprising to find 
that dust continuum emission from 
the brightest cluster galaxies in the most rapidly cooling clusters
being detected at sub-mm and MIR wavelengths (Edge et al. 1999, Egami et al. 2006, O'Dea et al. 2008).
The presence of cold molecular gas (Edge 2001, Salom\'e \& Combes 2003) and 
dust absorption in HST imaging (McNamara et al 1996) implies
that the dust continuum traces a substantial, cold 
component to the ISM in these massive elliptical galaxies.
However, the origin of the dust and how it is
shielded are still poorly understood.

The limitations with the current observations of dust emission
make it difficult to establish an unambiguous dust
mass as they do not sample over the peak of the dust
emission in the FIR.
The unprecedented sensitivity of {\it Herschel} \rm
(Pilbratt et al. 2010)
to FIR continuum offers the opportunity to 
accurately constrain the full FIR spectrum of
the dust emission in cluster cores.
 The authors were awarded 140~hours of time
in an Open Time Key Project (PI Edge) to investigate the FIR line and
continuum properties of a sample of 11 brightest cluster galaxies (BCGs)
in well-studied cooling flow clusters selected on the basis of optical
emission line and X-ray properties. 
The full goals of the 
project are to observe at least five atomic cooling lines for
each object that cover a range in density and temperature
behaviour and obtain a fully sampled FIR spectral energy
distribution.
In this paper we present
the Photodetector Array Camera \& Spectrometer (PACS, Poglitsch et al. 2010) and 
Spectral and Photometric Imaging REceiver (SPIRE, Griffin et al. 2010) 
photometry for the three targets observed in the Science Demonstration
Phase (SDP), Abell 1068 ($z=0.1386$), Abell 2597 ($z=0.0821$) and Zw3146 ($z=0.2906$). In
a parallel paper (Edge et al. 2010), we present the FIR spectroscopy for
the first two of these clusters. 

The three clusters observed have quite contrasting 
multiwavelength properties. Abell 1068 and Zw3146 both have 
strong MIR emission (O'Dea et al. 2008, Egami et al. 2006) with a relatively
bright CO detection (Edge 2001) and a weak central radio source
(McNamara et al. 2004). A1068 lies just below the luminosity 
threshold of a ULIRG (10$^{12}$~L$_\odot$) and exhibits
some contribution from an AGN (Crawford et al. 1999,
O'Dea et al. 2008). On the other hand, Abell 2597 
is a relatively weak MIR source (Donahue et al. 2007)
with a weak CO detection (Salome, priv. comm.) and
a powerful central radio source (Sarazin et al. 1995). The implied
FIR luminosity of A2597 is a factor of around 30 below that
of A1068 and, in addition, the fractional contribution from an AGN 
in the MIR is also lower in A2597.

\section{Observations}

We performed photometric imaging of A1068, A2597 and Zw3146 with PACS and SPIRE. 
The data were reduced with the {\it Herschel} Interactive Processing Environment~(HIPE) software version 2.3.1436  (Ott 2010). We used
for both PACS and SPIRE the official scripts as presented by the PACS and SPIRE ICC teams during the {\it Herschel} SDP 
data processing workshop in December 2009.

\subsection{PACS Data}

The PACS photometric 
observations were taken in LargeScanMapping mode in all three bands of the photometer, BS~(70~$\mu$m), BL~(100~$\mu$m) 
and R~(160~$\mu$m) using the medium scan speed  (20$''$s$^{-1}$).
The scan maps comprised 18 scan line legs of 4$'$
length and cross-scan step of 15$''$. Each observation had a ``scan'' and an orthogonal ``cross-scan'' direction and
we calibrated the corresponding data separately before combining them into a single map of  9$^{'} \times 9^{'}$. 
The resulting maps have a resolution of 5.2$''$, 7.7$''$ and 12$''$ at
70, 100 and 160~$\mu$m, respectively and are presented in the electronic version of this paper.
The PACS photometer performs dual-band imaging such that the BS and BL bands
each have simultaneous observations in the R band so we have two sets of scans in the R band.

We adopted the PACS Data Reduction Guideline to process the raw
level-0 data to calibrated level~2 products and used the official
script for PACS ScanMapping mode but with particular attention to the high pass filtering to remove ``1/$\sqrt{f}$'' noise. 
We choose to use the \textit{HighPassFilter} method with a filter of 20 readouts which will
remove structure on all scales above 82$''$.
The target BCG and other bright sources in the field were masked 
prior to applying the filter. The size of the mask was chosen to be less than the filter size so as to minimize any left-over low-frequency 
artefacts under the masks. We used masks with a radius of 15$''$ for our sources. We tried varying the size for the filter from 
10 to 30 readouts and the mask radius from 10--30$''$ and found our results to not change significantly for these ranges in values. 
Finally the task `photProject', was used to project the calibrated data onto a map on the sky in units of Jy pixel$^{-1}$. The 
``scan'' and ``cross-scan'' maps were then averaged to produce the final coadded map. The PACS and SPIRE images are included
in the electronic version of the paper.
The spatial flux distribution and flux densities of our target sources were investigated using cumulative flux curves. The spatial flux 
distribution for each of our three sources is consistent with that expected from a point source. Flux densities in the BS, BL and R 
band were extracted using a 33$\arcsec$ by 33$\arcsec$ aperture centered on the BCG. Small aperture corrections were applied as 
outlined in the PACS Scan Map release note (PICC-ME-TN-035). Care was taken to calibrate these derived flux densities to account 
for the known flux overestimation in the used HIPE version by factors 1.05, 1.09 and 1.29 in BS, BL and R bands respectively. 
The absolute flux accuracy is within 10~\% for BS and BL, and better than 20~\% for R. These uncertainties are not believed
to be correlated due to the BS and BL bands being taken at different times and the R band using a different detector.

\subsection{SPIRE Data}

The SPIRE photometry was performed in the LargeScanMap mode with cross-linked scans in two orthogonal scan directions. 
The photometer has a field of view of 4$^{'}\times 8^{'}$, which is observed simultaneously in three spectral bands, 
PSW~(250~$\mu$m), PMW~(350~$\mu$m) and PLW (500~$\mu$m) with a resolution of about 18$''$, 25$''$ 
and 36$''$, respectively. The resulting maps measure 12$^{'}\times 12^{'}$ in size and are presented in the electronic version
of this paper.

We used the standard HIPE pipeline for the LargeScanMap observing mode and the na\"ive map-maker.
The pre-processed raw telemetry data were first subject to 
engineering conversion wherein the raw timeline data were converted to meaningful units, the SPIRE pointing product was created, 
deglitching and temperature drift correction were performed, and maps were created, the units of which were Jy beam$^{-1}$.
Our targets are unresolved at the spatial resolution of SPIRE. We derived their flux densities by fitting the sources with the SPIRE point 
source response function. Care was taken to de-blend our target from other nearby sources at the longer
wavelengths, where the sources are most likely to be background to the cluster. We account for the known flux 
calibration offset in the used version of HIPE by applying the following multiplicative calibration factors 1.02, 1.05 and 0.94 to the 
derived flux densities in the PSW, PMW and PLW bands respectively (see Griffin et al. 2010, Swinyard et al. 2010).
We also performed
aperture photometry using the HIPE point-source extraction~(PSE) tool
but this method gives accurate results only for isolated point sources. At
350~$\mu$m and 500~$\mu$m, the BCGs in A2597 and Zw3146 are close to the
detection limit and at the confusion limit of SPIRE making the PSE method
of determining the fluxes unsuccessful. A1068 has a relatively strong
compact BCG in far infrared and so we performed the PSE to find that
the flux estimates using AIPS and HIPE agree with each other to better
than 5\%.

\begin{table}
\caption{Log of {\it Herschel}\rm \ Observations.}
\begin{tabular}{lllrcc}
\hline\hline
Cluster & $z$      & Instrument  &  $\lambda$ \ \ \  & Obsid  & Flux \\
        &          &              & ($\mu$m)   &        & (mJy)  \\
 & & & & &  \\
A1068   & 0.1386   & PACS    &  70 & 1342187051 &  542$\pm$6 \\
        &          & PACS    & 100 & 1342187053 &  757$\pm$6 \\
        &          & PACS    & 160 &           &  769$\pm$4 \\
        &          & SPIRE   & 250 & 1342187321 &  376$\pm$6 \\
        &          & SPIRE   & 350 &           &  135$\pm$6 \\
        &          & SPIRE   & 500 &           &   56$\pm$8 \\
        &          & SCUBA   & 450 &           &   39$\pm$13 \\
        &          & SCUBA   & 850 &           &  5.3$\pm$1.1 \\
        &          & {\it Spitzer} &  24 &           &  74.5$\pm$2.0 \\
        &          & {\it Spitzer} &  70 &           &  941$\pm$30 \\
        &          & {\it IRAS}    &  60 &           &  577$\pm$52 \\
        &          & {\it IRAS}    & 100 &           &  958$\pm$144 \\
A2597   & 0.0821   & PACS    &  70 & 1342187118 &   57$\pm$5 \\
        &          & PACS    & 100 & 1342187120 &   67$\pm$7 \\
        &          & PACS    & 160 &           &   86$\pm$4 \\
        &          & SPIRE   & 250 & 1342187329 &   30$\pm$6 \\
        &          & SPIRE   & 350 &           &   15$\pm$6 \\
        &          & SPIRE   & 500 &           &   $<16$ \\
        &          & SCUBA   & 850 &           &  14.5$\pm$2.3 \\
        &          & {\it Spitzer} &  24 &           &   2.1$\pm$0.2 \\
        &          & {\it Spitzer} &  70 &           &   49$\pm$6 \\
        &          & {\it Spitzer} & 160 &           &   52$\pm$3 \\
Zw3146  & 0.2906   & PACS    &  70 & 1342187043 &   94$\pm$6 \\
        &          & PACS    & 100 & 1342187045 &  150$\pm$6 \\
        &          & PACS    & 160 &           &  139$\pm$5 \\
        &          & SPIRE   & 250 & 1342187326 &   81$\pm$6 \\
        &          & SPIRE   & 350 &           &   30$\pm$6 \\
        &          & SPIRE   & 500 &           &  $<16$ \\
        &          & SCUBA   & 450 &           &  $<$48 \\
        &          & SCUBA   & 850 &           &  6.6$\pm$2.6 \\
        &          & {\it Spitzer} &  24 &           &   4.1$\pm$0.4 \\
        &          & {\it Spitzer} &  70 &           &   68$\pm$14 \\
        &          & {\it Spitzer} & 160 &           &  157$\pm$35 \\
\hline
 \end{tabular}
\tablefoot{The {\it Spitzer}
data are from Quillen et al. (2008), Donahue et al. (2007, priv. comm.) and Egami et al. (2006).
The SCUBA data are from Edge (priv. comm.), Zemcov et al. (2007) and Chapman et al. (2002).} 
\end{table}

\section{Results}

 In the PACS photometry,
A1068, A2597 and Zw3146 have been detected in all three bands. For A1068,  70 and 100$\mu$m 
values are slightly less than the {\it IRAS} 60 and 100 $\mu$m measurements. This could be due to nearby sources 
that cannot be separated from the BCG in the much lower resolution {\it IRAS} observations but no sufficiently
bright source is visible in our PACS imaging. 
There is a 
large difference between the {\it Spitzer} MIPS 70$\mu$m flux (Quillen et al. 2008) and our PACS 70$\mu$m 
flux, the PACS flux being a factor 1.7 lower than the MIPS flux. In the case of Zw3146 the MIPS and PACS 70 $\mu$m fluxes also differ 
with the PACS value being a factor 1.4 larger than the MIPS value (Egami et al. 2006).
For A2597 the PACS fluxes differ from the {\it Spitzer} 70 and 160 $\mu$m fluxes reported by Donahue et al. (2007). Part of this 
difference was resolved when the MIPS 70 $\mu$m data were re-analysed and found to be a factor of two too high 
(Donahue, priv. comm.). The differences observed between the PACS and {\it Spitzer} fluxes require further investigation.
In the SPIRE photometry, 
A1068 is detected in all three SPIRE bands. 
A2597 and Zw3146, while clearly detected in PSW and PMW bands, have a 1--2 $\sigma$ detection in the PLW band.
Table 1 gives the photometric results for the three galaxies, with $2\sigma$ upper-limit for A2597 and Zw3146 in PLW.
Figure 1 presents the radio to optical spectral energy distributions (SEDs) for the three targets. These plots show the 
significant variation in the relative radio-FIR-optical contributions for each of our galaxies. Here we focus on the sub-mm/MIR 
dust emission as sampled by PACS and SPIRE photometry, complemented by published {\it Spitzer} and {\it IRAS} measurements.

We fit the SEDs of the dust emission using black bodies modified
with a dust emissivity index, $\beta$.
The FIR-MIR slopes of our
sources require the presence of at least two dust
components. Previous studies of star-forming galaxies have 
indeed established that a single modified black body~(MBB) is
inadequate to account for the observed dust emission
(Wiklind 2003).  Hence, our model for the SEDs consists of two MBBs
with the dust emissivity index for each fixed to $\beta$=2 and
a mass absorption coefficient, $\kappa_{\rm d\nu}$, of 2.5~m$^2$~kg$^{-1}$ at 100$\mu$m.

For A1068 we fit the 24--850$\mu$m emission. For A2597 and Zw3146 the SCUBA 850$\mu$m detections have been removed and we fit 
only the 24--350$\mu$m range. In the case of A2597, this is due to the unknown amount of radio contamination at 850$\mu$m. In the case of 
Zw3146 the BCG is blended with strong background source at 850$\mu$m (Chapman et al. 2002). The
data are weighted in the fit inversely to the square of their error.
The resulting fits are shown in Figure 1.
The derived dust temperatures and total FIR luminosities for each source are listed in Table 2.

The results in Table~2 indicate that at least two dust components, one at 20--25 K and one at 50--60 K, are present in all three sources. 
The FIR emission is much stronger relative to the optical 
in A1068 and Zw3146 as compared to A2597. The SEDs of A1068 and Zw3146 resemble those of strongly star-forming systems and, based 
on the total FIR luminosity derived here, we find star formation rates (SFR) of 60 and 44 M$_{\odot}$ yr$^{-1}$ in these two systems
using the Kennicutt (1998) conversion factor.
For A2597 a much more modest SFR of 2 M$_{\odot}$ yr$^{-1}$ is inferred. These values are comparable to SFRs
derived from H$\alpha$ line and/or UV continuum emission given the uncertainties of these tracers.
However, the SFR values derived from {\it Spitzer} data are higher for A1068 and Zw3146. The
difference for A1068 is the most pronounced and can be directly attributed to the stronger AGN contribution in this
object (Quillen et al. 2008) which boosts the 24$\mu$ flux compared other comparable sources. Therefore, when the
total FIR luminosity is derived from the  15$\mu$m flux infered from {\it Spitzer}
it will be overestimated. The value for Zw3146 from
Egami et al. (2006) is higher than ours as their fit includes the SCUBA 850$\mu$m point from Chapman et al. (2002)
which appears to be overestimated on the basis of our SPIRE data.

The gas to dust ratio is found to be between 80 and 140 (see Table 2). Gas temperatures 
can be inferred from CO measurements (Edge 2001, Salome \& Combes 2003). These estimates infer gas 
temperatures of 25--40~K thus implying that the gas and dust share a common environment and are 
potentially co-located in the denser regions of cold, molecular gas clouds. 
We have attempted to determine how much extended
emission is present from our highest spatial
resolution PACS 70$\mu$m image but we find no
evidence for more than 10\% additional flux
beyond a point source. Clearly these limits
will improve with a better characterisation of
the instrument but we believe that we can 
conclude that the dust emission in our targets
has an extent comparable to that the bulk of the CO
emitting gas and optical emission lines ($<5''$
or 5--20~kpc).

\begin{table}
\caption{Summary of results and other cluster properties.}
\begin{tabular}{lccc}
\hline\hline
Cluster &   A1068    &    A2597   &   Zw3146   \\
        &            &            & \\
Dust Temperatures & 24$\pm$4K        & 21$\pm$6K        & 23$\pm$5K \\
                  & 57$^{+12}_{-4}$K & 48$^{+17}_{-5}$K & 53$^{+22}_{-6}$K \\
Cold Dust Mass & 5.1$\times 10^{8}$~M$_\odot$ &  2.3$\times 10^{7}$~M$_\odot$ &  5.4$\times 10^{8}$~M$_\odot$ \\
Warm Dust Mass & 3.9$\times 10^{6}$~M$_\odot$ &  2.9$\times 10^{5}$~M$_\odot$ &  1.9$\times 10^{6}$~M$_\odot$ \\
Total FIR Luminosity  &   3.5$\times 10^{11}$~L$_\odot$ &  8.8$\times 10^{9}$~L$_\odot$ &  2.5$\times 10^{11}$~L$_\odot$ \\
Star Formation Rate &  60$\pm$20~M$_\odot$~yr$^{-1}$  &  2$\pm$1~M$_\odot$~yr$^{-1}$  & 44$\pm$14~M$_\odot$~yr$^{-1}$ \\
SFR {\it Spitzer}   &  188~M$_\odot$~yr$^{-1}$        &  4~M$_\odot$~yr$^{-1}$        & 70$\pm$14~M$_\odot$~yr$^{-1}$ \\
SFR {\it optical/UV} &  20--70~M$_\odot$~yr$^{-1}$     &  10--15~M$_\odot$~yr$^{-1}$   & 47$\pm$5~M$_\odot$~yr$^{-1}$ \\
CO gas mass & 4.1$\times 10^{10}$~M$_\odot$ &   2.0$\times 10^{9}$~M$_\odot$  &  7.7$\times 10^{10}$~M$_\odot$ \\
H$\alpha$ Slit Luminosity & 8$\times 10^{41}$~erg~s$^{-1}$ & 3$\times 10^{41}$~erg~s$^{-1}$ & 3$\times 10^{42}$~erg~s$^{-1}$ \\
\hline
 \end{tabular}
\tablefoot{The {\it Spitzer} SFR values are from O'Dea et al. (2008), Donahue et al. (2007)
and Egami et al. (2006). The Optical/UV SFR values are from
McNamara et al. (2004), Donahue et al. (2007) and Egami et al. (2006).
The CO gas masses are from Edge (2001) and Salom\'e (priv. comm.) and
the  H$\alpha$ slit luminosities are from Crawford et al. (1999).}
\end{table}

\section{Discussion and conclusions}

Our initial {\it Herschel} results confirm the presence of
the striking dust emission peak expected from the
observations at sub-mm (Edge et al. 1999, Chapman et al. 2002)
and MIR (Egami et al. 2006, O'Dea et al. 2008). 

The star formation rates derived from the full-sampled
FIR SED are comparable to those derived from {\it Spitzer}
24$\mu$m fluxes apart from  A1068, which has the strongest
contribution from an AGN so  
hot dust dominates to the 24$\mu$m flux.
However, in the sub-mm the contribution from the
radio continuum from an active nucleus must be
correctly accounted for before any dust mass can
be estimated from the 850$\mu$m flux. In the case of
A2597 here and A2390 in Edge et al. (1999), the
presence of a powerful radio source appears to
contribute to the SCUBA 850$\mu$m flux. 

While it is difficult to draw any general 
conclusions from just three BCGs, we note
with interest that the ratio of dust mass to
CO-derived gas mass is constistent for all
three within a factor of five. If the 
dust were mostly generated through dust
ejection from evolved stars then
the dust mass should closely correlate with the
total stellar mass. However, our three galaxies
have very similar optical/NIR absolute magnitudes.
So, unless the ejected dust were ``captured'' 
by the cold gas clouds  protecting it from
X-ray sputtering, this suggests
that the apparent correlation between the molecular gas
and dust masses arises from a direct connection between
the gas reservoir and star formation.

These results are a limited example of those
to come in the very near future from {\it Herschel} \rm
as there are two other Open Time Key Projects
(PI Egami and Smith) that are targetting 
a total of 70 clusters that cover a broad range of
BCG properties so the wider context of these
initial observations can be determined.
In particular, the amount of dust present
in more quiescent BCGs and other massive
cluster ellipticals will be important
in assessing how much of the 
dust seen in cool core BCGs orginates
from the underlying stellar population.

\begin{figure}
\centerline{\includegraphics[width=5.7cm,angle=90]{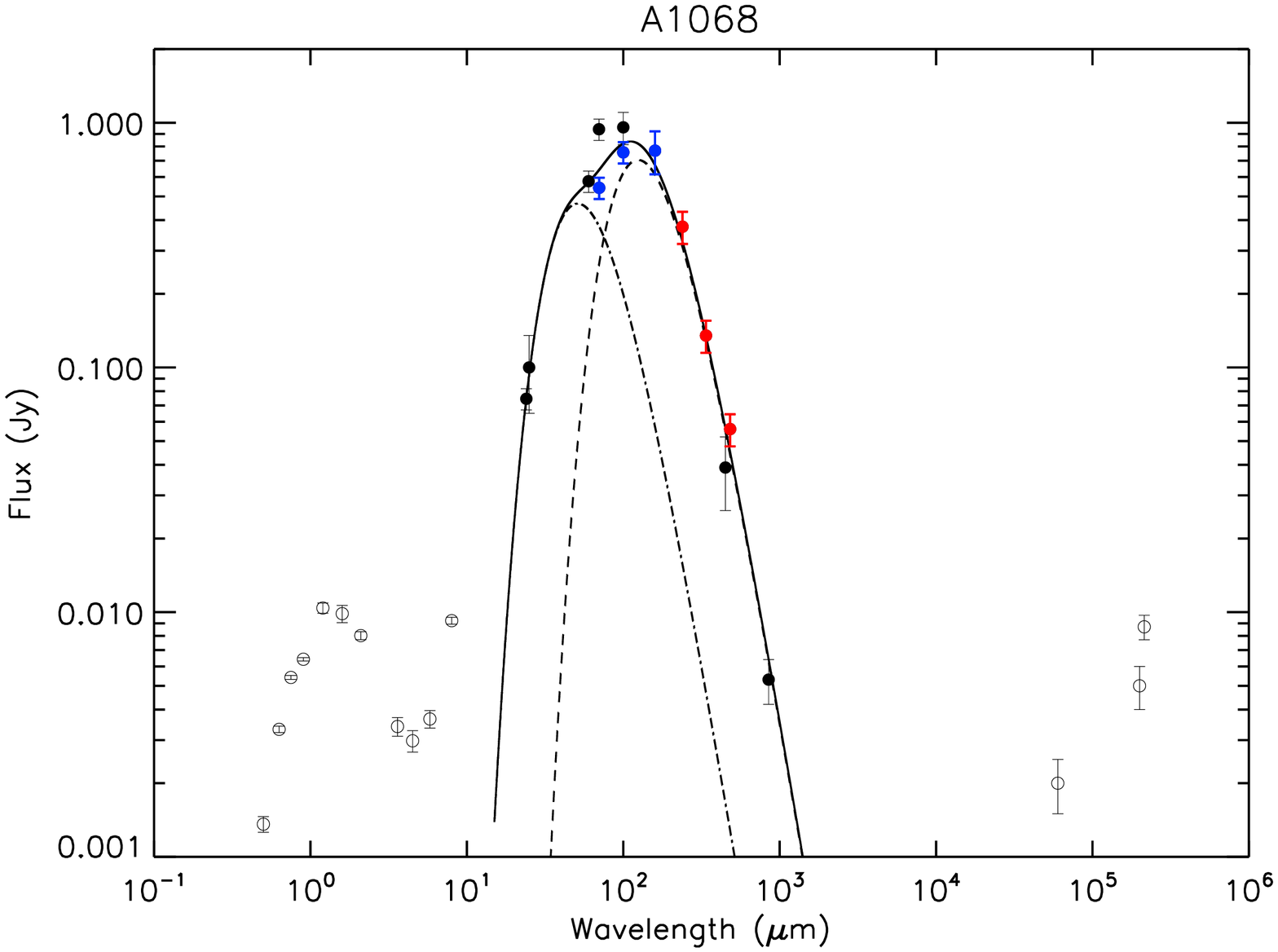}}
\centerline{\includegraphics[width=5.7cm,angle=90]{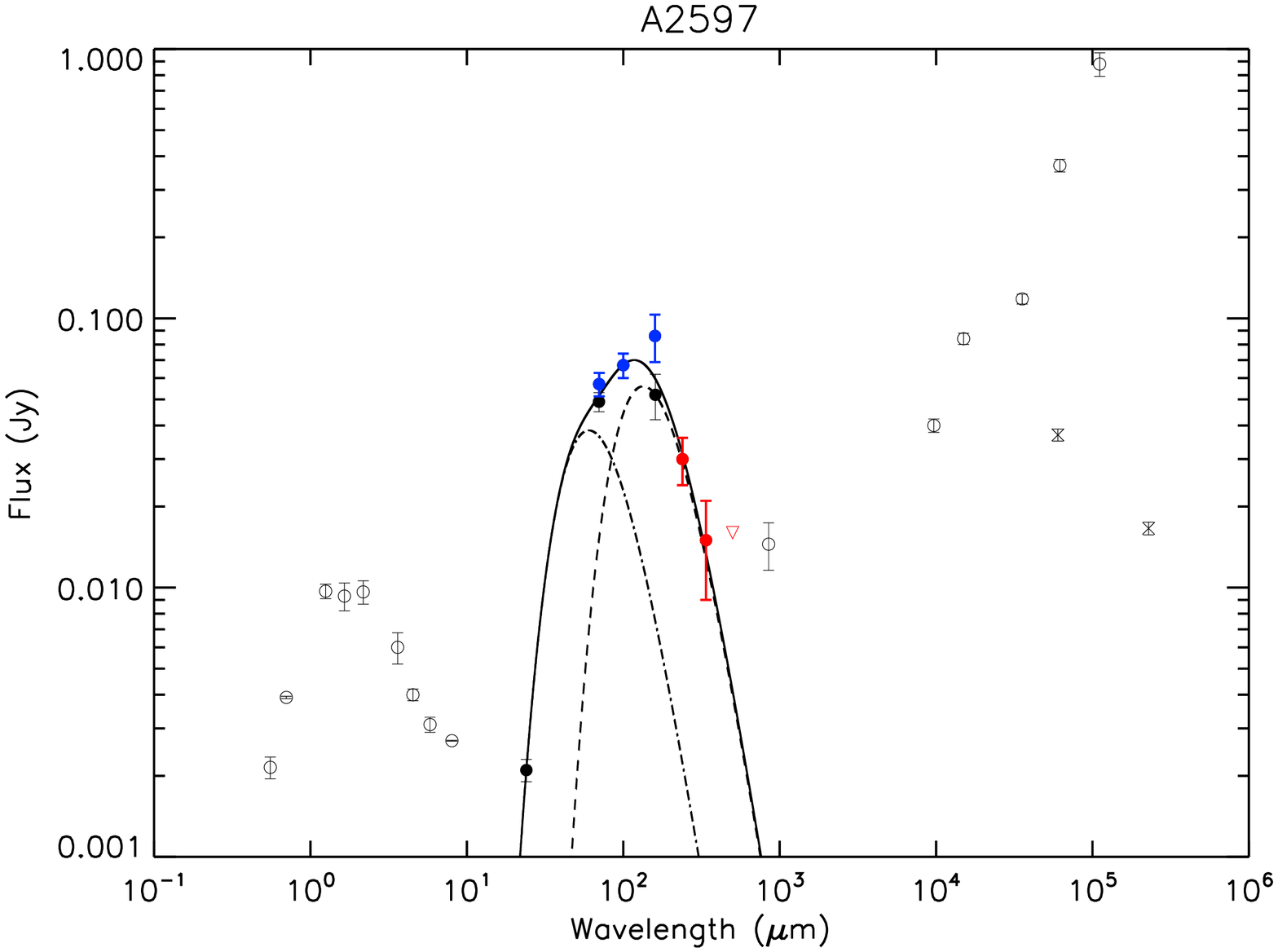}}
\centerline{\includegraphics[width=5.7cm,angle=90]{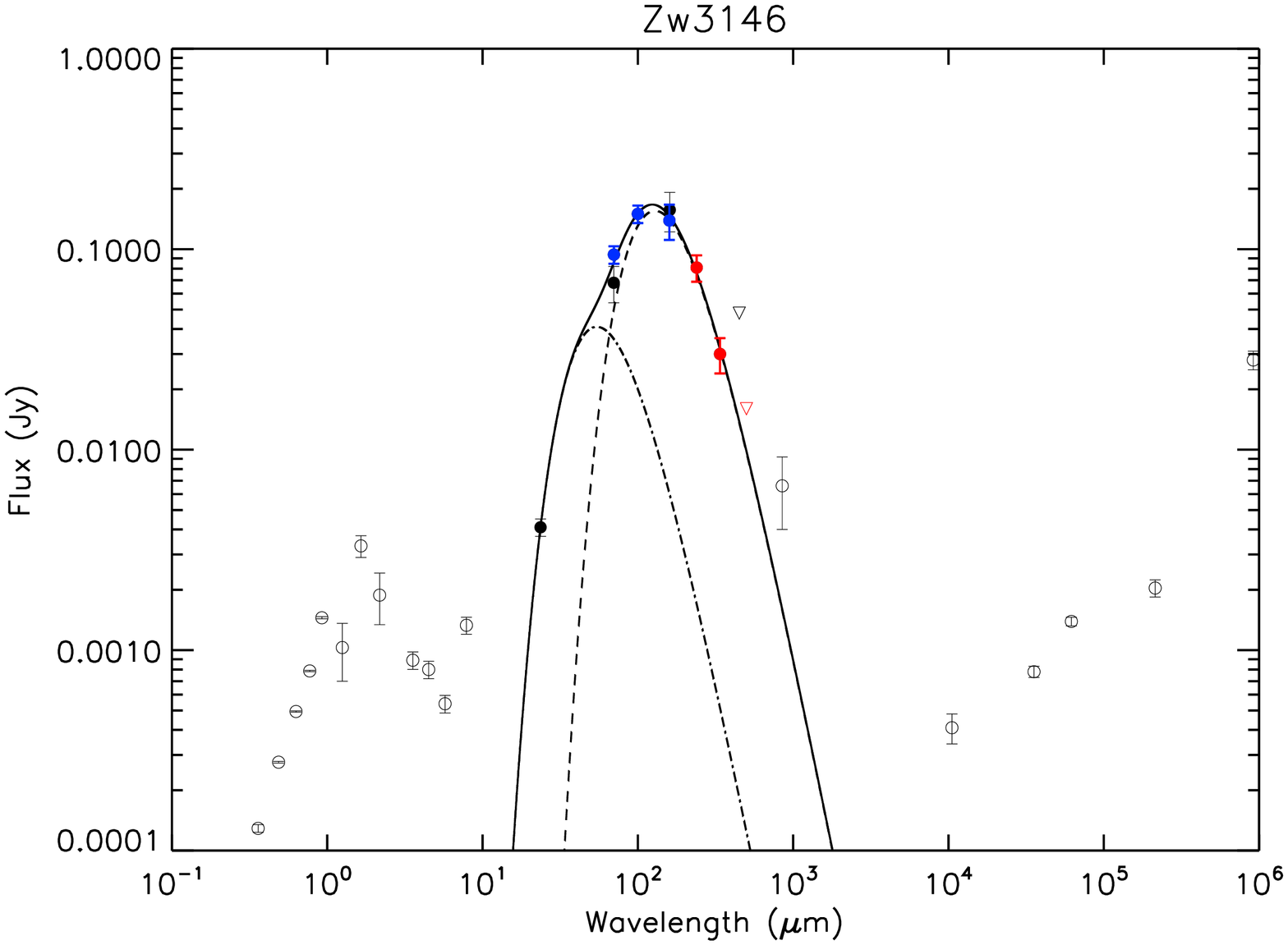}}
\caption{
Spectral energy distributions for A1068 (top), A2597 (middle) and Zw3146 (bottom) including 
{\it Herschel} \ PACS/SPIRE (blue/red symbols), {\it Spitzer}, Radio, NIR photometry from 2MASS 
and optical photometry from SDSS. 
To account for absolute flux 
uncertainties we have set the following errors on the fluxes derived from the various 
instruments (unless the quoted error is larger than this); PACS BS/BL 10\%, PACS R 20\%, SPIRE 15\%, 
{\it Spitzer} 10\%, SCUBA 450 $\mu$m 30\% and SCUBA 850 $\mu$m 20\%. The model fit to the 
sub-mm/FIR/MIR data is shown by the black solid line. Only filled symbols have been used 
in the fit. The two modified blackbodies making up the model are shown by the black 
long dash and dash-dot lines. For A2597 we also show two VLBI measurements (black 
crosses) of the BCG core at 1.3 and 5 GHz (Taylor et al. 1999). These points show that
 the BCG has a strong, inverted radio core.}
\end{figure}

\begin{acknowledgements}
We would like to thank the {\it Herschel} Observatory and instrument teams for the extraordinary 
dedication they have shown to deliver such a powerful telescope. 
We would like to thank 
the HSC and NHSC consortium for help with data reduction pipelines. J.B.R.O. thanks HSC, 
the {\it Herschel} Helpdesk and the PACS group at MPE for useful discussions. R.\,M. thanks the NHSC for the HIPE tutorials.
\end{acknowledgements}

\newpage

\begin{figure*}
\centerline{\includegraphics[width=16.6cm]{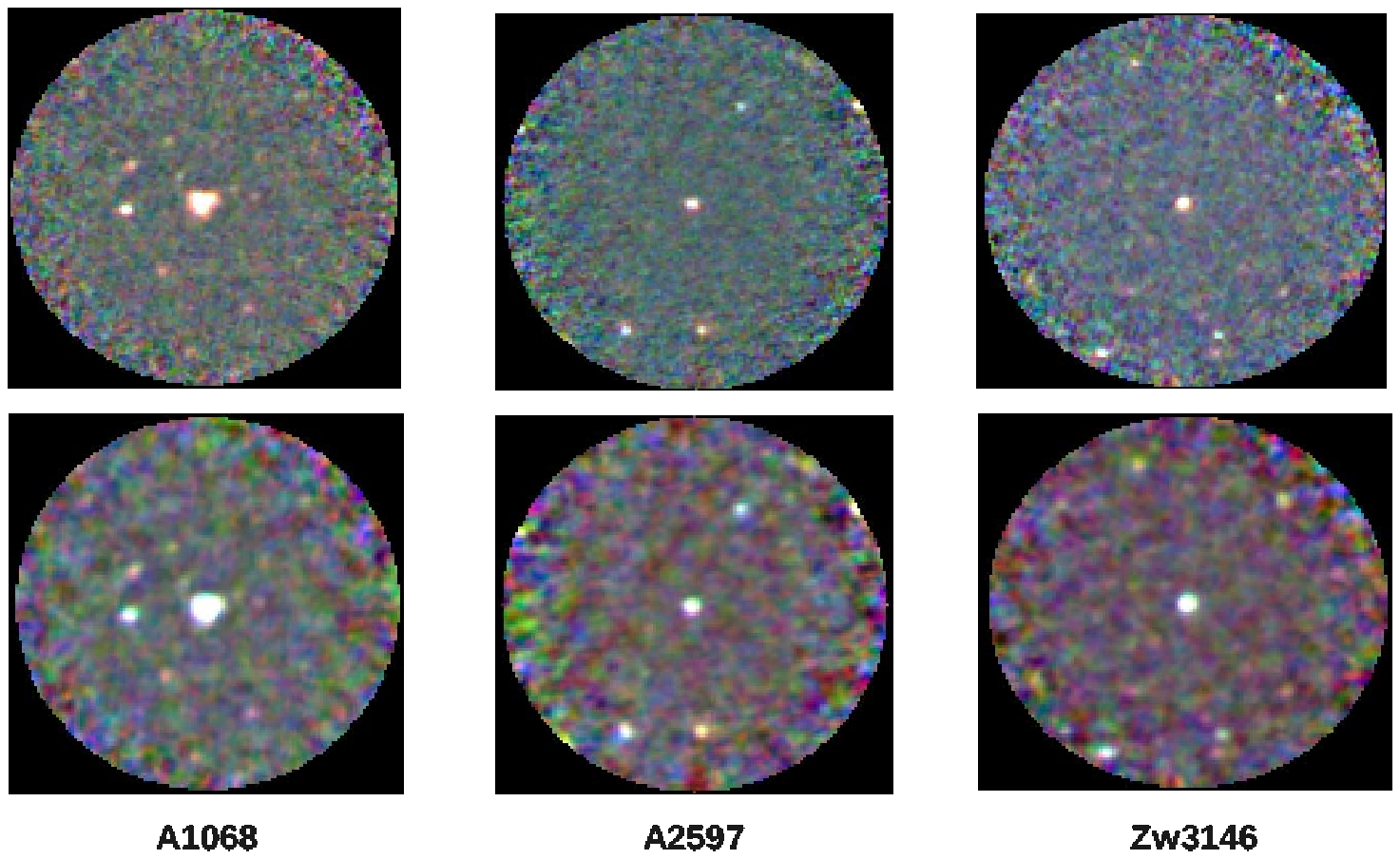}}
\caption{
Colour images from the three PACS bands (BS, BL and R in the blue, green
and red channels) for the three clusters within radius of 2.5$'$ of the BCG
The top row are images combined
in their original  resolution and the bottom row are the images combined
with a common smoothing of 12$''$ to match resolution.
}
\end{figure*}

\begin{figure*}
\centerline{\includegraphics[width=16.6cm]{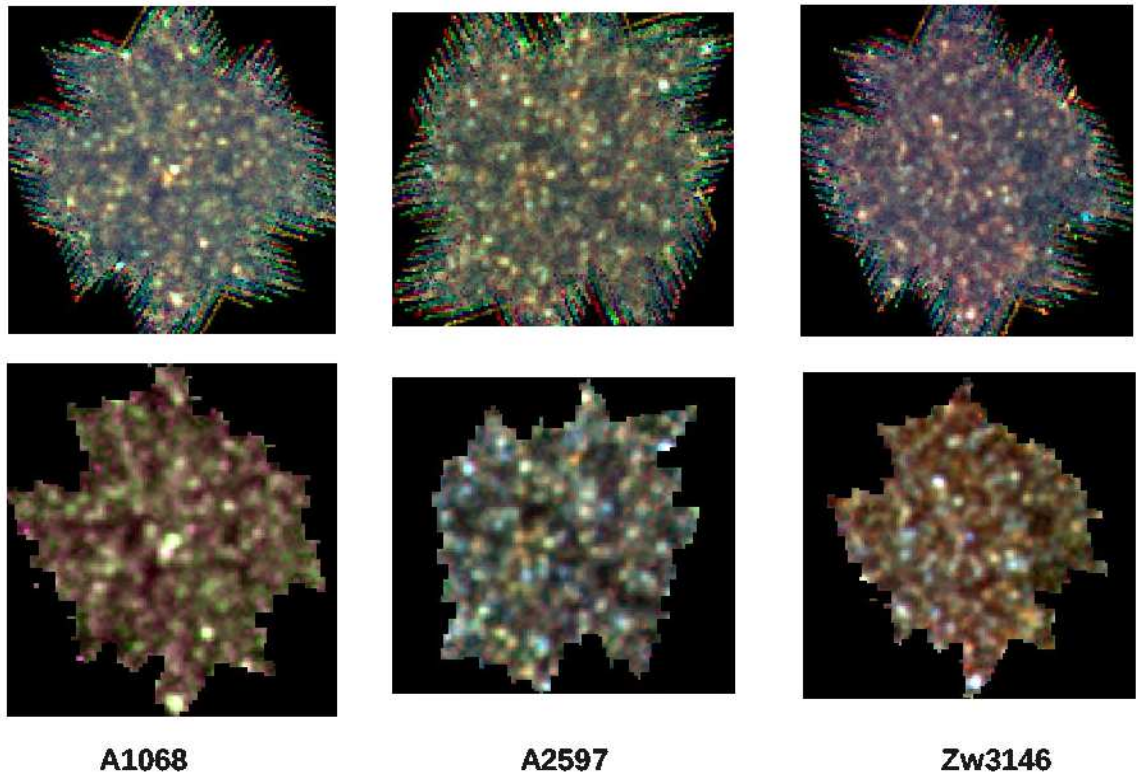}}
\caption{
Colour images from the three SPIRE bands (PSW, PMW and PLW in the blue, green
and red channels) for full field covered for the three clusters covering
approximately 12$' \times 12'$.
The top row are images combined
in their original  resolution and the bottom row are the images combined
with a common smoothing of 36$''$ to match resolution and clipped to
remove areas of low exposure. The BCG is at the centre of the image
and in A2597 and Zw3146 is the bluest object present (see text).
} 
\end{figure*}

\end{document}